---------------------------------------------------------------------------

\documentstyle[12pt]{article}
\title{ Levinson's Theorem for Non-local Interactions in Two Dimensions}

\author{Shi-Hai Dong\thanks{Electronic address: DONGSH@BEPC4.IHEP.AC.CN}\\
{\scriptsize Institute of High Energy Physics, P. O. Box 918(4), 
Beijing 100039, The People's Republic of China}\\
\\
Xi-Wen Hou\\
{\scriptsize Institute of High Energy Physics, 
P. O. Box 918(4), Beijing 100039}\\
{\scriptsize and Department of Physics, University of Three
Gorges, Yichang 443000, The People's Republic of China}\\
\\
Zhong-Qi Ma\\
{\scriptsize China Center for Advanced Science and Technology
(World Laboratory), P. O. Box 8730, Beijing 100080}\\
{\scriptsize  and Institute of High Energy Physics, P. O. Box 918(4) 
Beijing 100039, The People's Republic of China}}

\date{}

\begin{document}

\maketitle

\begin{abstract}
In the light of the Sturm-Liouville theorem, the Levinson theorem for 
the Schr\"{o}dinger equation with both local and non-local cylindrically
symmetric potentials is studied. It is proved that the two-dimensional 
Levinson theorem holds for the case with both local and non-local 
cylindrically symmetric cutoff potentials, which is not necessarily 
separable. In addition, the problems related to the positive-energy 
bound states and the physically redundant state are also discussed 
in this paper.


\end{abstract}

\newpage

\begin{center}
{\large 1. Introduction}\\
\end {center}

The Levinson theorem [1], an important theorem in the scattering theory, 
established the relation between the total number of bound states and the
phase shifts at zero momentum. The Levinson theorem has been proved 
by several authors with different methods, and generalized 
to different fields [2-9]. 
Generally speaking, three main methods are used 
to prove the Levinson theorem. One [1] is based on the elaborate analysis 
of the Jost function. This method requires nice behavior of the potential. 
The second, based on the Green function method [5], expounds that 
the total number of the physical states, which is infinite, is proved 
to be independent of the potential and the number of the 
bound states is the difference between the infinite numbers
of the scattering states without and with the potential. 
The third method proves the Levinson theorem by the Sturm-Liouville 
theorem [6-8]. This simple, intuitive method is easy to be 
generalized. Some obstacles and ambiguities, which may occur 
in other two methods, disappear in the third method. We have 
succeeded in dealing with the nonrelativistic and relativistic 
problems in two dimensions by this way [10-11]. 

The reasons why we write this paper are that, on the one hand
the Levinson theorem in two dimensions has been studied 
in experiment [12] as well as in theory [13,10,11] because of
the wide interest in lower-dimensional field theories and 
other modern physics [14-20], on the other hand the Levinson 
theorem for non-local interactions in two
dimensions has never appeared in the literature.

This paper is organized as follows. In Sec. 2, we establish the 
Sturm-Liouville theorem for non-local interactions in two dimensions. 
The Levinson theorem for this case will be set up in Sec. 3. Some 
problems related to the positive-energy bound states and the physically 
redundant state will be studied in Sec. 4 and 5.

\vskip 0.5cm
\begin{center}
{\large 2. The Sturm-Liouville Theorem}
\end{center}

Throughout this paper $\hbar=1$ and the mass $\mu=1/2$ are employed 
for simplicity. Consider the Schr\"{o}dinger equation with a local
potential $V(r)$ and a non-local potential $U(r,r')$, where both 
potentials are cylindrically symmetric
$$\begin{array}{rl}
H\psi(r,\varphi)
&=~-\left(\displaystyle {1 \over r}\displaystyle {\partial
\over \partial r} r \displaystyle {\partial \over \partial r} 
+\displaystyle {1 \over r^{2}} \displaystyle {\partial^{2}
\over \partial \varphi^{2} } \right) \psi(r,\varphi) +
V(r) \psi(r,\varphi) \\
&~~~+\displaystyle \int U(r,r') \delta(\varphi'-\varphi)
\psi(r',\varphi') r'dr'd\varphi  \\
&=~E \psi(r,\varphi). \end{array} \eqno (1) $$

The mesonic theory of nuclear forces points out the interaction 
between two nucleons is local at great distances but becomes 
non-local if the two nucleons come close. To simplify the 
expression, we assume, following Martin [4], that the non-local 
potential $U(r,r')$ is real, symmetric, vanishing at large 
distance, and not too singular at the origin [21], 
$$\begin{array}{l}
U(r,r')=U(r',r) \\
|r^{2}U(r,r')|\sim 0,~~~~~{\rm at}~~r\sim 0, \\
U(r,r')=0,~~~~~~~~~~~~ {\rm when}~~~~r\geq r_{0}. 
\end{array} \eqno (2) $$

\noindent
As usual, the local potential $V(r)$ should not be too singular
at the origin and at infinity. We assume that $V(r)$ satisfies 
$$\begin{array}{ll}
|r^{2}V(r)|\sim 0,~~~~~&{\rm at}~~r\sim 0, \\
V(r)=0,~~~~~~~~~~~~ &{\rm when}~~~~r\geq r_{0}.
\end{array} \eqno (3) $$

\noindent
The first condition is necessary for the nice behavior of the 
wave function at the origin [1], and the potential with the second 
condition is called a cutoff one, namely, it is vanishing beyond 
a sufficiently large radius $r_{0}$. It was proved that the tail 
of the local potential at infinity will not change the essence 
of the proof [10] if it decays faster than $r^{-3}$ at infinity. 
Under this assumption, the integral range in (1) is, in fact, 
from $0$ to $r_{0}$, and the equation in the region $[r_{0},\infty)$ 
becomes that for the free particle.

Introduce a parameter $\lambda$ for the potentials
$$V(r,\lambda)=\lambda V(r),~~~~~U(r,r',\lambda)=\lambda U(r,r').
\eqno (4) $$

\noindent
As $\lambda$ increases from zero to one, the potentials
$V(r,\lambda)$ and $U(r,r',\lambda)$ change from zero
to the given potentials $V(r)$ and $U(r,r')$, respectively.

Owing to the symmetry of the potentials, letting
$$\psi(r,\phi,\lambda)=r^{-1/2}R_{Em}(r,\lambda)e^{\pm im\phi},
~~~~~~~~m=0,1,2,\ldots, \eqno (5) $$

\noindent
we obtain the radial equation
$$\begin{array}{l}
\displaystyle {\frac{\partial^2}{\partial r^2}}
R_{Em}(r,\lambda)+\left(E-V(r,\lambda)-\displaystyle 
{\frac{m^2-1/4}{r^2}}\right)R_{Em}(r,\lambda)\\
~~~=\sqrt{r} \displaystyle \int U(r,r',\lambda)R_{Em}(r',\lambda)\sqrt{r'} dr', 
\end{array} \eqno (6)$$

\noindent
where $\pm m$ and $E$ denote the angular momentum and energy, 
respectively. Since the radial function $R_{Em}(r,\lambda)$ is 
independent of the sign of the angular momentum, we only 
discuss the case with non-negative $m$. 

We are going to solve the radial equation (6) in two regions
$[0,r_{0}]$ and $[r_{0},\infty)$, and match two solutions at $r_{0}$.
Since the Schr\"{o}dinger equation is linear, the wavefunction
can be multiplied by a constant factor. Removing the effect
of the factor, we only need one matching condition at $r_{0}$
for the logarithmic derivative of the radial function:
$$A_{m}(E,\lambda)\equiv 
\left\{\displaystyle {1 \over R_{Em}(r,\lambda)}
\displaystyle {\partial R_{Em}(r,\lambda)\over \partial r}
\right\}_{r=r_{0}-}
=\left\{\displaystyle {1 \over R_{Em}(r,\lambda)}
\displaystyle {\partial R_{Em}(r,\lambda)\over \partial r}
\right\}_{r=r_{0}+}. \eqno (7) $$

We now turn to the Sturm-Liouville theorem. Denote by 
$\overline{R}_{Em}(r,\lambda)$ a solution of (6) with the 
energy $\overline{E}$. Multiplying the equations for 
$R_{Em}(r,\lambda)$ and $\overline{R}_{Em}(r,\lambda)$ by 
$\overline{R}_{Em}(r,\lambda)$ and $R_{Em}(r,\lambda)$, 
respectively, and calculating their difference, we have
$$\begin{array}{l}
\displaystyle {\partial \over \partial r}\left(
R_{Em}(r,\lambda)\displaystyle {\partial \overline{R}_{Em}(r,\lambda)
\over \partial r}
-\overline{R}_{Em}(r,\lambda)\displaystyle {\partial R_{Em}(r,\lambda)
\over \partial r} \right)
+(\overline{E}-E)R_{Em}(r,\lambda)\overline{R}_{Em}(r,\lambda)\\
~~~=\sqrt{r} R_{Em}(r,\lambda)\displaystyle \int U(r,r',\lambda)
\overline{R}_{Em}(r',\lambda) \sqrt{r'} dr' \\
~~~~~-\sqrt{r} \overline{R}_{Em}(r,\lambda)
\displaystyle \int U(r,r',\lambda)
R_{Em}(r',\lambda)\sqrt{r'} dr'. \end{array} \eqno (8) $$

\noindent
According to the boundary condition, both $R_{Em}(r,\lambda)$ 
and $\overline{R}_{Em}(r,\lambda)$ go to zero when $r$ tends
to zero. Integrating (8) over the variable $r$ in the region 
$[0, r_{0}]$ and noting the symmetric property of $U(r,r')$, 
we have
$$\begin{array}{l}
\displaystyle {1 \over \overline{E}-E} \left\{
R_{Em}(r,\lambda)\displaystyle {\partial \overline{R}_{Em}(r,\lambda)
\over \partial r}
-\overline{R}_{Em}(r,\lambda)\displaystyle {\partial R_{Em}(r,\lambda)
\over \partial r} \right\}_{r=r_{o}-}\\
~~~=-\displaystyle \int_{0}^{r_{0}}R_{Em}(r',\lambda)
\overline{R}_{Em}(r',\lambda)dr'.
\end{array} $$

\noindent
Taking the limit, we obtain
$$\begin{array}{rl}
\displaystyle {\partial A_{m}(E,\lambda) \over \partial E}&\equiv~ 
\displaystyle {\partial \over \partial E} \left(
\displaystyle {1 \over R_{Em}(r,\lambda)}
\displaystyle {\partial R_{Em}(r,\lambda)\over \partial r}
\right)_{r=r_{0}-} \\
&=~-R_{Em}(r_{0},\lambda)^{-2}
\displaystyle \int_{0}^{r_{0}}R_{Em}(r',\lambda)^{2} dr'<0.
\end{array} \eqno (9) $$

\noindent
Similarly, from the boundary condition that when $E\leq 0$ 
the radial function $R_{Em}(r,\lambda)$ tends to zero at infinity,
we have
$$\displaystyle {\partial \over \partial E} \left(
\displaystyle {1 \over R_{Em}(r,\lambda)}
\displaystyle {\partial R_{Em}(r,\lambda)\over \partial r}
\right)_{r=r_{0}+}
=R_{Em}(r_{0},\lambda)^{-2}
\displaystyle \int_{r_{0}}^{\infty}R_{Em}(r',\lambda)^{2} dr'>0.
 \eqno (10) $$

\noindent
Therefore, when $E\leq 0$, both sides of the matching condition
(7) are monotonic with respect to the energy $E$. As energy
increases, the logarithmic derivative of the radial function at 
$r_{0}-$ decreases monotonically, but that at $r_{0}+$
increases monotonically. This is an expression for the
Sturm-Liouville theorem [22].

\vskip 0.5cm
\begin{center}
{\large 3. The Levinson Theorem}\\
\end{center}

The establishment of the Levinson theorem for the case with 
both local and non-local cylindrically symmetric potentials 
is similar to that for the case with only a local potential.

In solving the radial equation (6) in the region $[0,r_{0}]$, 
only one solution is convergent at the origin. Thus, for the 
given potentials the logarithmic derivative $A_{m}(E,\lambda)$ 
is determined in principle. For example, for free particle 
($\lambda=0$) we have
$$R_{Em}(r,0)=\left\{\begin{array}{ll}
\sqrt{\displaystyle  {\pi kr \over 2 }}J_{m}(kr),~~~~~  
&{\rm when}~~E>0~~{\rm and}~~k=\sqrt{E} \\
e^{-im\pi/2}\sqrt{\displaystyle {\pi \kappa r \over 2 }}J_{m}(i\kappa r)
,~~~~~&{\rm when}~~E\leq 0~~{\rm and}~~\kappa=\sqrt{-E}, 
\end{array} \right. \eqno (11) $$

\noindent
where the factor in front of the radial function $R_{Em}(r)$ 
is not important. The solution $R_{Em}(r,0)$ given in (11) 
is a real function. The logarithmic derivative at $r_{0}-$ 
for $E\leq 0$ is
$$A_{m}(E,0)\equiv 
\left\{\displaystyle {1 \over R_{Em}(r,0)}
\displaystyle {\partial R_{Em}(r,0)\over \partial r}
\right\}_{r=r_{0}-}
=\left\{\begin{array}{ll}
\displaystyle {k J'_{m}(k r_{0}) 
\over J_{m}(k r_{0}) }-\displaystyle {1 \over 2r_{0}}
&{\rm when}~~E>0 \\
\displaystyle {i\kappa J'_{m}(i\kappa r_{0}) 
\over J_{m}(i\kappa r_{0}) }-\displaystyle {1 \over 2r_{0}}
&{\rm when}~~E\leq 0 
\end{array} \right. \eqno (12) $$

In the region $[r_{0},\infty)$, we have $V(r)=U(r,r')=0$. For $E>0$, 
there are two oscillatory solutions to (6). Their combination 
can always satisfy the matching condition (7), so that there
is a continuous spectrum for $E>0$.
$$\begin{array}{rl}
R_{Em}(r,\lambda)&=~\sqrt{\displaystyle  {\pi kr \over 2 }}
\left\{ \cos \eta_{m}(k,\lambda)J_{m}(kr)
-\sin \eta_{m}(k,\lambda) N_{m}(kr)\right\}\\
&\sim~ \cos \left(kr-\displaystyle {m\pi \over 2}-
\displaystyle  {\pi \over 4} +\eta_{m}(k,\lambda) \right) ,~~~~~~~~~~
{\rm when}~~r\longrightarrow \infty , \end{array} \eqno (13) $$

\noindent
where $N_{m}(kr)$ is the Neumann function. The phase shift 
$\eta_{m}(k,\lambda)$ is determined by the matching condition (7)
$$\tan \eta_{m}(k,\lambda)=\displaystyle {J_{m}(kr_{0}) \over
N_{m}(kr_{0})} ~\cdot ~\displaystyle {A_{m}(E,\lambda)
-kJ'_{m}(kr_{0})/J_{m}(kr_{0})-1/2r_{0}
\over A_{m}(E,\lambda)-kN'_{m}(kr_{0})/N_{m}(kr_{0})-1/2r_{0} } ,
 \eqno (14) $$
$$\eta_{m}(k)\equiv \eta_{m}(k,1) , \eqno (15) $$

\noindent
where the prime denotes the derivative of the Bessel function, 
the Neumann function, and later the Hankel function with respect 
to their argument. Although the radial equation (6) in the
region $[r_{0},\infty)$ is independent of $\lambda$, the solution
$R_{Em}(r,\lambda)$ and the phase shift $\eta_{m}(k,\lambda)$
do depend on $\lambda$ through the matching condition (7).

The phase shift $\eta_{m}(k,\lambda)$ is determined from (14)
up to a multiple of $\pi$ due to the period of the tangent function.
Levinson determined the phase shift $\eta_{m}(k)$ with respect
to the phase shift $\eta_{m}(\infty)$ at the infinite momentum. 
For any finite potential, the phase shift $\eta_{m}(\infty)$
will not change and is always equal zero. Therefore, Levinson's 
definition for the phase shift is equivalent to the convention that 
the phase shift $\eta_{m}(k,\lambda)$ is determined with respect to 
the phase shift $\eta_{m}(k,0)$ for the free particle, where 
$\eta_{m}(k,0)$ is defined to be zero
$$\eta_{m}(k,0)=0,~~~~{\rm where}~~\lambda=0. \eqno (16) $$

\noindent
There is some ambiguity for $\eta_{m}(\infty)$ when a bound
state with a positive energy occurs (see Sec. 4). However, 
as far as the Levinson theorem is concerned, the latter
convention is more convenient. We prefer to use this convention
where the phase shift $\eta_{m}(k)$ is determined
completely as $\lambda$ increases from zero to one. It is 
the reason why we introduce the parameter $\lambda$.

For $E\leq 0$ there is only one convergent solution at infinity
$$R_{Em}(r)=e^{i(m+1)\pi/2}\sqrt{\displaystyle {\pi \kappa r \over 2 }}
H^{(1)}_{m}(i\kappa r) \sim e^{-\kappa r} ,~~~~~
{\rm when}~~r\longrightarrow \infty . \eqno (17) $$

\noindent
where $H^{(1)}_{m}(x)$ is the Hankel function of the first kind.
Thus, the matching condition (7) is not always satisfied. 
When the matching condition (7) is satisfied, a bound state 
appears at this energy. It means that there is a discrete 
spectrum for $E \leq 0$.

\noindent
From (17) we have
$$\begin{array}{rl}
\left\{\displaystyle {1 \over R_{Em}(r,0)}
\displaystyle {\partial R_{Em}(r,0)\over \partial r}
\right\}_{r=r_{0}+}
&=~\displaystyle {i\kappa H^{(1)}_{m}(i\kappa r_{0})' 
\over H^{(1)}_{m}(i\kappa r_{0}) }-\displaystyle  {1 \over 2r_{0}}\\
&=~\left\{\begin{array}{ll} (-m+1/2)/r_{0}\equiv \rho_{m} 
&{\rm when}~~E\longrightarrow 0 \\
-\kappa \sim -\infty &{\rm when}~~E\longrightarrow  -\infty.
\end{array} \right. \end{array} \eqno (18) $$

On the other hand, if $V(r)=U(r,r')=0$, we obtain from (12)
$$\begin{array}{rl}
A_{m}(E,0)&\equiv~ 
\left\{\displaystyle {1 \over R_{Em}(r,0)}
\displaystyle {\partial R_{Em}(r,0)\over \partial r}
\right\}_{r=r_{0}-}\\
&=~\displaystyle {i\kappa J'_{m}(i\kappa r_{0}) 
\over J_{m}(i\kappa r_{0}) }-\displaystyle {1 \over 2r_{0}}
=\left\{\begin{array}{ll} (m+1/2)/r_{0} &{\rm when}~~E\longrightarrow 0 \\
\kappa \sim \infty &{\rm when}~~E\longrightarrow  -\infty.
\end{array} \right. \end{array} \eqno (19) $$

\noindent
It is easy to see from (18) and (19) that as energy increases 
from $-\infty$ to $0$, there is no overlap between two variant 
ranges of the logarithmic derivatives at two sides of $r_{0}$
such that there is no bound state when $\lambda=0$ except for 
$S$ wave where there is a half bound state at $E=0$. The half 
bound state will be discussed at the end of this section.

If $A_{m}(0,\lambda)$ decreases across the value $\rho_{m}\equiv
(-m+1/2)/r_{0}$ as $\lambda$ increases, an overlap between two 
variant ranges of the logarithmic derivatives at two sides of 
$r_{0}$ appears. Since the logarithmic derivatives of the radial 
function at $r_{0}-$ decreases monotonically as the energy increases, and
that at $r_{0}+$ increases monotonically, the overlap means that there must
be one and only one energy where the matching condition (7) is satisfied,
i.e. a scattering state changes to a bound state.

As $\lambda$ increases, a zero point in the zero energy solution
$R_{0m}(r,\lambda)$ may comes through $r_{0}$. In this process
$A_{m}(0,\lambda)$ may decrease to the negative infinity, jumps 
to the positive infinity, and decreases again, or vice versa.
It is not a singularity. If $A_{m}(0,\lambda)$ decreases, through
the jump at infinity, again across the value $\rho_{m}$, 
another bound state appears. 

As $\lambda$ increases from zero to one, each time $A_{m}(0,\lambda)$ 
decreases across the value $\rho_{m}$, a new overlap between 
the variant ranges of two logarithmic derivatives appears such that 
a scattering state changes to a bound state. Conversely, each time 
$A_{m}(0,\lambda)$ increases across the value $\rho_{m}$, an 
overlap between those two variant ranges disappears 
such that a bound state changes back to a scattering state.
The number of bound states $n_{m}$ is equal to the times that 
$A_{m}(0)$ decreases across the value $\rho_{m}$ as 
$\lambda$ change from zero to one, subtracted by the times that $A_{m}(0)$ 
increases across the value $\rho_{m}$. In the following, we will 
show from (14) that this number is nothing but the phase shift 
$\eta_{m}(0)$ at zero momentum divided by $\pi$.

It is easy to see from (14) that the phase shift $\eta_{m}(k,\lambda)$
increases monotonically as the logarithmic derivative $A_{m}(E)$
decreases
$$\left. \displaystyle  {\partial \eta_{m}(k,\lambda) \over 
\partial A_{m}(E,\lambda)}
\right|_{k}=-\displaystyle  {8r_{0}\cos^{2}\eta_{m}(k) \over
\pi \left(2r_{0}A_{m}(E)N_{m}(kr_{0})-2kr_{0}N'_{m}(kr_{0})-
N_{m}(kr_{0})\right)^{2} } \leq 0. \eqno (20) $$

\noindent
The phase shift $\eta_{m}(0,\lambda)$ is the limit of the phase shift 
$\eta_{m}(k,\lambda)$ as $k$ tends to zero. Therefore, what we are 
interested in is the phase shift $\eta_{m}(k,\lambda)$ at sufficiently
small momentum $k$, $k\ll 1/r_{0}$. For the small momentum $k$
we obtain from (14)

$$\begin{array}{l}
\tan \eta_{m}(k,\lambda)\sim \\[2mm]
\sim\left\{\begin{array}{ll}
\displaystyle  {-\pi (kr_{0})^{2m} \over 2^{2m}m!(m-1)!}
~\cdot~\displaystyle  {A_{m}(0,\lambda)-(m+1/2)/r_{0} \over 
A_{m}(0,\lambda)-c^{2}k^{2}-\rho_{m} \left(1-
\displaystyle  {(kr_{0})^{2} \over (m-1)(2m-1) }\right) } 
&{\rm when}~~m \geq 2 \\
\displaystyle  {-\pi (kr_{0})^{2} \over 4}~\cdot~
\displaystyle  {A_{m}(0,\lambda)-3/(2r_{0}) \over 
A_{m}(0,\lambda)-c^{2}k^{2}-\rho_{1}
\left(1+ 2(kr_{0})^{2}\log(kr_{0})\right) } 
&{\rm when}~~m = 1 \\
\displaystyle  {\pi \over 2\log(kr_{0})}~\cdot~
\displaystyle  {A_{m}(0,\lambda)-c^{2}k^{2}-\rho_{0} \left(
1-(kr_{0})^{2} \right) \over
A_{m}(0,\lambda)-c^{2}k^{2}-\rho_{0}\left(1+
\displaystyle  {2 \over \log (kr_{0})} \right) }
 &{\rm when}~~m =0 .\end{array}\right. \end{array} \eqno (21) $$

\noindent
In addition to the leading terms, we include in (21) 
some next leading terms, which is useful only for the 
critical case where the leading terms are canceled with each other.

First of all, it can be seen from (21) that $\tan \eta_{m}(k,\lambda)$
tends to zero as $k$ goes to zero, i.e. $\eta_{m}(0,\lambda)$ is
always equal to the multiple of $\pi$. In other words, if the
phase shift $\eta_{m}(k,\lambda)$ for a sufficiently small $k$ 
is expressed as a positive or negative acute angle plus $n\pi$, 
where $n$ is an integer, its limit $\eta_{m}(0,\lambda)$
is equal to $n\pi$. It means that $\eta_{m}(0,\lambda)$ changes 
discontinuously. By the way, in three dimensions, 
the phase shift at zero momentum of $S$ wave may have an additional
$\pi/2$ when the half bound state occurs. 

Secondly, since the phase shift $\eta_{m}(k,\lambda)$
increases monotonically as the logarithmic derivative $A_{m}(E,\lambda)$
decreases, the phase shift at zero momentum $\eta_{m}(0,\lambda)$
will jump by $\pi$ if $\tan \eta_{m}(k,\lambda)$ at the sufficiently small
$k$ changes sign from positive to negative as $A_{m}(E,\lambda)$ decreases, 
and vice versa. When $\lambda$ changes from zero to one continuously, 
each time $A_{m}(0,\lambda)$ decreases from near and larger than the 
value $\rho_{m}$ to smaller than that value, the denominator in 
(21) changes sign from positive to negative and the remaining factor 
keeps positive, such that the phase shift at zero momentum 
$\eta_{m}(0,\lambda)$ jumps by $\pi$. Conversely, each time 
$A_{m}(0,\lambda)$ increases across the value $\rho_{m}$, the 
phase shift at zero momentum $\eta_{m}(0,\lambda)$ jumps by 
$-\pi$. Therefore, the phase shift $\eta_{m}(0)/\pi$ 
is just equal to the times $A_{m}(0,\lambda)$ decreases across the 
value $\rho_{m}$ as $\lambda$ increases from zero to one, 
subtracted by the times $A_{m}(0,\lambda)$ increases across that value.
Therefore, we proved the Levinson theorem for the Schr\"{o}dinger
equation in two dimensions for non-critical cases:
$$\eta_{m}(0)=n_{m}\pi. \eqno (22a) $$

Thirdly, we should pay some attention to the case of $m=0$. When 
$A_{0}(0,\lambda)$ decreases across the value $\rho_{0}=1/(2r_{0})$, both 
the numerator and denominator in (21) change signs, but not 
spontaneously because the next leading terms in the numerator and 
denominator in (21) are different. It is easy to see that the 
numerator changes sign first, and then the denominator changes 
sign, i.e. $\tan \eta_{0}(k,\lambda)$
at small $k$ changes firstly from negative to positive, then
to negative again such that $\eta_{0}(0,\lambda)$ jumps by $\pi$.
Similarly, when $A_{m}(0,\lambda)$ increases across the value
$\rho_{0}$, $\eta_{0}(0,\lambda)$ jumps by $-\pi$.

When $\lambda=0$ and $m=0$, the numerator in (21) is equal 
to zero, the denominator is positive, and the phase shift 
$\eta_{0}(0)$ is defined to be zero. If $A_{0}(E)$ decreases
when $\lambda$ increases from zero, the numerator
becomes negative firstly, and then the denominator changes from
positive to negative such that the phase shift $\eta_{0}(0)$
jumps by $\pi$ and simultaneously a new bound state appears. 

Finally, we turn to discuss the critical cases where a half bound
state occurs. If the logarithmic derivative $A_{m}(0,1)$ 
is equal to the value $\rho_{m}$, the following solution with
zero energy in the region $[r_{0},\infty)$ will match this 
$A_{m}(0,1)$ at $r_{0}$
$$R_{0m}(r,1)=r^{-m+1/2}. $$

\noindent
It is a bound state when $m\geq 2$, but called a half bound state
when $m=1$, and $0$. A half bound state is a zero energy solution
of the Schr\"{o}dinger equation which is finite but does not decay 
fast enough at infinity to be square integrable.
We are going to discuss the critical case where $A_{m}(0,\lambda)$ 
decreases (or increases) and reaches, but not across, 
the value $\rho_{m}$ as $\lambda$ increases from a value, a 
little smaller than one, to one. For definiteness, we discuss 
the case where $A_{m}(0,\lambda)$ decreases and reaches the value 
$\rho_{m}$ as $\lambda$ increases to one. In this case 
a new bound state with zero energy appears for $m \geq 2$, but 
does not appear for $m=1$ and $0$. We need to check whether or 
not the phase shift $\eta_{m}(0)$ increases an additional $\pi$.

It is evident to find that the denominator in (21) 
for $m \geq 2$ has changed the sign 
from positive to negative as $A_{m}(0)$ decreases and reaches the value
$\rho_{m}$, i.e. the phase shift $\eta_{m}(0)$ jumps by
$\pi$ and simultaneously a new bound state of zero energy
appears. 

For $m=0$ the next leading term of the denominator in (21) is positive
and larger than the term $-c^{2}k^{2}$, such that
the denominator does not change sign, i.e. the phase shift
$\eta_{m}(0)$ does not jump. It meets the fact that no new
bound state appears. 

For $m=1$ the next leading term of the denominator in (21) 
is negative such that the denominator does change sign, i.e. 
the phase shift $\eta_{m}(0)$ jumps by $\pi$ as $A_{m}(0)$ 
decreases and reaches the value $\rho_{1}$. However, in this 
case no new bound state appears simultaneously. 

The discussion for the cases where $A_{m}(0)$ increases and 
reaches the value $\rho_{m}$ is similar. Therefore, 
Levinson's theorem (22a) holds for the critical cases except 
for $m=1$. In the latter case, Levinson's theorem for the 
Schr\"{o}dinger equation with both local and non-local
interactions in two dimensions becomes:
$$\eta_{m}(0)=\left(n_{m}+1 \right)\pi, ~~~~~
{\rm when}~~m=1, ~~{\rm and~a~ half~ bound~ state~ occurs}. \eqno (22b) $$

As discussed above, it is found that the Levinson theorem 
holds without any modification for the case where a non-local 
potential is included.

\vskip 0.5cm
\begin{center}
{\large 4. Positive-energy bound states}\\
\end{center}

It is well known that, in the case with only a local interaction, 
the wavefunction and its first derivative would never vanish at 
the same point except for the origin, so there is no positive-energy 
bound state. However, in the case with a non-local interaction, 
Martin showed that the solution with an asymptotic form is not 
unique when the potential satisfies some conditions [4], i.e. 
there exists the positive-energy bound state with a vanishing 
asymptotic form. If a small perturbative potential is added
such that the non-local potential satisfies the conditions, the 
positive-energy bound state will appear and the phase shift at 
this energy increases rapidly by almost $\pi$. This can be seen 
explicitly in the examples given by Martin [4] and Kermode [23].

It was pointed out by Kermode that the inverse tangent function is not
single-valued and it is physically more satisfactory to include a jump by
$\pi$ to the phase shift at the energy $E_{0}$, where a positive-energy 
bound state occurs. Martin and Chadan [4,21] defined the
phase shift to be continuous even at $E_{0}$ so that an additional $\pi$
will be included into $\delta(0)-\delta(\infty)$ for each
positive-energy bound state. This is their reason to modify Levinson's
theorem by the term $\sigma\pi$ where $\sigma$ denotes the number of 
positive-energy bound state. But according to the viewpoint of Kermode, no
modification to the Levinson theorem is required.
 
However, the phase shift at zero energy in our convention doesn't 
change, no matter which viewpoint is used, i.e. no matter whether 
the phase shift jumps or not at the energy with a positive-energy 
bound state. Therefore, the Levinson theorem (22) holds for the 
cases where positive-energy bound states may occur.

\vskip 0.5cm
\begin{center}
{\large 5. Redundant state}\\
\end{center}

The resonating group model of the scattering of nuclei, or 
other composite systems, derives an effective two-body 
interaction in which a non-local potential appears. There 
are some physically redundant states which describe Pauli-forbidden 
states for the compound system, and the physical two-body states 
must be orthogonal to these redundant states [24]. In the case 
of three dimensions, Saito [25], Okai, ${\rm et~al}$ 
[26] and Englefield-Shoukry [27] proposed a simple non-local 
term which guarantees that the required orthogonality, and
verified that it was a good representation of the interactions. 
If there is just one redundant state represented by the real 
normalized wavefunction $U(r)$, then the two-dimensional 
Saito's equation is
$$\begin{array}{l}
\displaystyle {\frac{d^2}{dr^2}}R_{Em}(r)+\left(E-V(r)
-\displaystyle {\frac{m^2-1/4}{r^2}}\right)R_{Em}(r)\\
~~~=U(r)\displaystyle \int_{0}^{\infty}U(s)\left(\displaystyle 
{\frac{d^2}{ds^2}}-V(s)-\displaystyle {\frac{m^2-1/4}
{s^2}}\right)R_{Em}(s)  ds, \\
\displaystyle \int_{0}^{\infty} U^{2}(s)ds =1 ,\end{array} \eqno (23) $$

\noindent
and
$$E \int_{0}^{\infty}U(r)R_{Em}(r) dr=0. \eqno (24)$$

\noindent
The solution of (23) satisfies the orthogonality constraint 
except for that of zero energy. Saito's non-local potential 
is separable.

If the Schr\"{o}dinger equation with only a local potential
$V(r)$ has a bound state with a negative $-{\cal E}<0$, the 
corresponding wavefunction is denoted by $\psi(r)$:
$$\begin{array}{l}
\displaystyle {\frac{d^2}{dr^2}}\psi(r)-\left(V(r)
+\displaystyle {\frac{m^2-1/4}{r^2}}\right)
\psi(r)={\cal E}\psi(r),\\
\displaystyle \int_{0}^{\infty}\psi(r)^{2} dr=1.
\end{array} \eqno (25)$$

It is obvious that $U(r)=\psi(r)$ satisfies (23) with zero
energy. Therefore, it is the so-called physically redundant 
state. As far as a mathematical equation (23) is concerned, 
the redundant state is one of the bound states with zero energy.

\vspace{10mm}
{\bf Acknowledgments}. This work was supported by the National
Natural Science Foundation of China and Grant No. LWTZ-1298 of
the Chinese Academy of Sciences. 



\end{document}